\documentclass[showpacs,superscriptaddress,aps,twocolumn,epsfig,rotate,rotating,pre]{revtex4}
\usepackage[final]{graphicx}

\begin{document}

\title{Topological Effects caused by the Fractal Substrate on the Nonequilibrium
Critical Behavior of the Ising Magnet}

\author{M. A. Bab}
\email{mbab@inifta.unlp.edu.ar} 
\affiliation{Instituto de Investigaciones Fisicoqu\'{i}micas Te\'{o}ricas y Aplicadas, Facultad de Ciencias Exactas,
Universidad Nacional de La Plata, Sucursal 4, CC 16 (1900) La Plata, Argentina.}
\affiliation{Departamento de F\'{i}sica, Facultad de Ciencias Exactas, Universidad Nacional de La Plata, 
La Plata, Argentina}

\author{G. Fabricius} 
\email{fabriciu@fisica.unlp.edu.ar}
\affiliation{Instituto de Investigaciones Fisicoqu\'{i}micas Te\'{o}ricas y Aplicadas, Facultad de Ciencias Exactas,
Universidad Nacional de La Plata, Sucursal 4, CC 16 (1900) La Plata, Argentina.}
\affiliation{Research Associate of the Abdus Salam International Centre for Theoretical Physics, Trieste, Italy}

\author{E. V. Albano}
\email{ealbano@inifta.unlp.edu.ar} 
\affiliation{Instituto de Investigaciones Fisicoqu\'{i}micas Te\'{o}ricas y Aplicadas, Facultad de Ciencias Exactas,
Universidad Nacional de La Plata, Sucursal 4, CC 16 (1900) La Plata, Argentina.}

\begin{abstract}
The nonequilibrium critical dynamics of the Ising magnet on a fractal
substrate, namely the Sierpinski carpet with Hausdorff dimension $d_H$
=1.7925, has been studied within the short-time regime by means of Monte
Carlo simulations. The evolution of the physical observables was followed at
criticality, after both annealing ordered spin configurations (ground state)
and quenching disordered initial configurations (high temperature state),
for three segmentation steps of the fractal. The topological effects become
evident from the emergence of a logarithmic periodic oscillation
superimposed to a power law in the decay of the magnetization and its
logarithmic derivative and also from the dependence of the critical
exponents on the segmentation step. These oscillations are discussed in the
framework of the discrete scale invariance of the substrate and carefully
characterized in order to determine the critical temperature of the
second-order phase transition and the critical exponents corresponding to
the short-time regime. The exponent $\theta $ of the initial increase in the
magnetization was also obtained and the results suggest that it would be
almost independent of the fractal dimension of the susbstrate, provided that 
$d_H$ is close enough to $d=2$.
\end{abstract}
\pacs{05.45.Df, 64.60.Ht, 75.10.Hk, 02.70.Uu}
\maketitle

\section{Introduction}

The study and characterization of continuous phase transitions occurring on
fractal substrates have attracted much attention during the past decades in
the research fields of statistical physics and materials science. The very
well-known general belief in universality states that for a given symmetry
of the order parameter and range of interactions, the critical behavior only
depends on the dimensionality d, i.e. the influence of the underlying
structure becomes negligible at the critical point when the correlation
length is much larger than the cell spacing. In particular, Ising models
embedded in different lattices in the absence of external magnetic fields,
with short-range interactions and for a given d $>$1, exhibit a continuous
phase transition characterized by identical critical exponents, with the
critical temperature decreasing to zero at the lower critical dimension d=1.
Instead of the replication of an elementary cell by translation, fractals
are constructed by the iteration of a generating cell, consequently the
topological details of the generating cell are present at any scale. In this
way, the critical exponents also depend on other geometric and topological
parameters, such as the ramification and lacunarity of the fractal\cite
{Hao87,Monceau04}. It has been shown that a second-order phase transition at
nonzero temperature occurs only if the fractal substrate has an infinite
ramification order\cite{Gefen83,Gefen84,Gefen80}. On the other hand, the
lacunarity, which was introduced by Mandelbrot in order to give a measure of
the deviation from the translational invariance \cite{Mandelbrot}, can be
obtained considering the departure from the power law of the mass scaling
definition of the fractal Hausdorff dimension (d$_H$), namely $m(R)\propto
R^{d_H}$ (where $m(R)$ is the mass inside a sphere of radius R centered
inside a fractal structure). Direct quantitative studies of the Ising model
on fractal substrates have shown that the values of the ratio of critical
exponents $\gamma /\nu $ depend upon the lacunarity and indicate that by
increasing the lacunarity one has to introduce higher corrections in the
finite-size scaling associated with the critical behavior of physical
observables\cite{Monceau04}.
Most of the previous work on this topic was based on the same class of
fractals, built in the same way as the standard Sierpinski carpet (SC(b,c)),
which have an infinite ramification order. A hypercube in d dimensions is
segmented into b$^d$ subhypercubes and c$^d$ of them are then removed, this
segmentation process is iterated on the remaining subhypercubes. The
mathematical fractal is obtained after an infinite number of segmentation
steps (k). However, for practical purposes, a fractal is constructed by
applying a finite number of segmentation steps. So, the physical observables
show not only finite-size effects caused by the size of the system, but also
a lower cutoff due to the segmentation step. As a matter of fact, results
obtained by performing Monte Carlo simulations depend on the number of
segmentation steps, so it is interesting to know the way these observables
converge to the ones expected for the mathematical fractal. The key point
here is that in order to obtain critical exponents from simulations in
equilibrium, one has to apply finite-size scaling analysis that is hindered
by topological scaling corrections. These corrections and the critical
slowing down increase when the fractal dimension decreases towards one.
Furthermore, due to these effects the segmentation steps studied could not
be large enough to truly observe the asymptotic behavior. So, under these
circumstances it is not surprising that also the critical exponents could
not be easily obtained\cite{Monceau01,Monceau04}. Furthermore, Pruessner et
al \cite{Pruessner} have questioned the validity of finite-size scaling on
fractal substrates arguing that each segmentation step represents a new
thermodynamic system that cannot be treated as a scaled version of the
previous one.

During the last years, nonequilibrium critical dynamics has been developed
and a dynamic scaling form was found that is already valid in the short-time
regime \cite{Janssen,Zheng98}. In addition to the dynamic exponent z, the
static exponents, originally defined in equilibrium, also appear in the
short-time scaling form. This fact leads to new methods for the numerical
measurement of all the static and dynamic critical exponents, as well as the
critical temperature. Since the measurements now are carried out within the
short-time regime, they do not suffer from critical slowing down. Because of
the small nonequilibrium correlation length, it is also easy to overcome
finite-size effects. Due to the above-mentioned advantages the
nonequilibrium critical dynamics within the short-time regime is a promissory
tool for the study of continuous phase transitions on fractal substrates.
Hitherto, the topological effects on the short-time dynamic behavior and
the possible corrections to scaling have not been systematically
investigated. Recently, we have reported soft oscillatory deviations from
the power-law decay of the magnetization in the short-time regime\cite{Bab05}
for the Ising model on the SC(3,1) with d$_H$ = 1.8927. On the other hand,
logarithmic periodic oscillatory deviations in the behavior of physical
observables have been reported for several systems that present fractal
characteristics embedded into the dynamics of the model and/or the
substrate. Some examples are the wave propagation in fractal systems \cite
{Sornette}, the Blume-Capel model on the Sierpinski gasket\cite{Lessa}, the
dynamics of biological systems such as the bronchial tree\cite{Shlesinger},
proteins \cite{Shen}, magnetic and resistive effects on a system of wires
connected along the Sierpinski gasket\cite{Sornette}, and random walk
through fractal environments\cite{Isliker}.

\smallskip The purpose of this paper is to study the occurrence of
topological effects on the scaling laws that describe the nonequilibrium
short-time critical dynamics of a magnetic system on a fractal substrate.
For this purpose we have performed extensive numerical Monte Carlo
simulations of the Ising model on the SC(4,2). Three different segmentation
steps are investigated in order to show how the lower cutoff due to the
finite-segmentation step affects both the critical exponents and critical
temperature. This procedure also allows us to study the convergence of the
measured values to the values expected for the mathematical fractal.

\smallskip In Section II we introduce the model, we give the simulation
details and remark on the main features of the short-time dynamic scaling. In
Section III we present, analyze and discuss the short-time dynamic
behavior of systems starting from both disordered (III A) and ordered
initial states (III B). Critical exponents obtained under these two
conditions are compared. Also, in subsection III C we discuss the
topological characteristics of the substrate and their influence on the
dynamic critical behavior of the physical observables. In Section VI we
present our conclusions.

\section{The Model and the Methods}

\subsection{The Ising Model on Sierpinski Carpets}

\begin{figure}[ht]
\includegraphics[height=11cm,width=8cm,angle=-90]{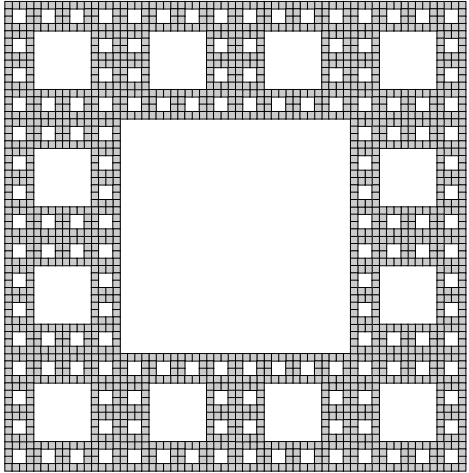} 
\caption{Sketch of the Sierpinski carpet SC(4,2) iterated up to the $k=3$
segmentation step. Spins are placed at the center of filled squares.}
\label{fig1}
\end{figure}

Figure \ref{fig1} shows a sketch of the fractal substrate used in the
present work, namely the SC(4,2), for the segmentation step k = 3. The
Hausdorff dimension of the mathematical fractal is given by $d_H=\frac{\log
(4^2-2^2)}{\log 4}\simeq $1.7925. In the case of the fractal, the convergence of
relevant properties towards the thermodynamic limit occurs when the
structure is constructed by the iterative process. Such convergence is
reflected in the dependence of the mean number of nearest neighbors per site
on the segmentation step, which converges to a constant value as k goes to
infinity. In order to study the critical behavior by means of Monte Carlo
simulations it is necessary to simulate large enough segmentation steps to
ensure that the asymptotic region is reached. For the SC(4,2) the deviation
of the mean number of next-nearest neighbors from the value corresponding to
the thermodynamic limit (mathematical fractal), using periodic boundary
conditions and determined by means of the transfer-matrix method, becomes
negligible for k $\geq $ 5 \cite{Monceau02}.

As mentioned, the studied fractal has an infinite ramification order, which
implies that the Ising model would exhibit a continuous phase transition at
finite temperature. In order to perform the simulation the spins were placed
at the center of the occupied subsquares and consequently, the number of
spins increases as a power law of the lattice size with an exponent given by 
$d_H$. The Hamiltonian of the system is then given by

\begin{equation}
H=-J\sum_{\left\langle i,j\right\rangle }s_i\;s_j,  \label{ec1}
\end{equation}
\noindent  where $s_i$ assumes the values $\pm 1$, the sum runs over all
interacting first-neighbor pairs and the positive exchange coupling constant 
$J$ corresponds to the ferromagnetic case.

\subsection{Short-Time Dynamic Behavior at Second-Order Phase Transitions}

In this section we give a brief summary on the short-time dynamic (STD)
method for characterization of continuous phase transitions. According to
STD arguments\cite{Janssen}, a magnetic system at high temperature and with
a small magnetization $m_0\ll $1 suddenly quenched to the critical
temperature T$_c$ presents a universal dynamical evolution, which sets right
after a microscopic time scale t$_{mic},$ large enough in the microscopic
sense but still very small in the macroscopic sense. If during this time
interval the nonequilibrium correlation length remains smaller than the
system size the short-time dynamics is free of finite-size effects.
According to the short-time dynamic scaling\cite{Zheng98,Zheng99} the
magnetization, its second moment and the autocorrelation function should
follow power-law scaling relations given by

\begin{equation}
M(t)=\left\langle \frac 1N\sum_{i=1}^Ns_i\right\rangle \sim m_0\;t^\theta
F\left( t^{\theta +\frac \beta {\nu z}}\,m_0\right) ,  \label{ec2}
\end{equation}

\begin{equation}
M^2(t)=\left\langle \left( \frac 1N\sum_{i=1}^Ns_i\right) ^2\right\rangle
\propto t^{\left[ \frac{d_{eff}}z-\frac{2\beta }{\nu z}\right] },
\label{ec3}
\end{equation}

\begin{equation}
A(t)=\left\langle \frac 1N\sum_{i=1}^Ns_i(t)\,\,s_i(0)\right\rangle \propto
t^{-\lambda },\,\,\,\text{with }\lambda =(\frac{d_{eff}}z-\theta ),
\label{ec4}
\end{equation}
\noindent  respectively. Here $\left\langle ...\right\rangle $ denotes the
average over samples, $N$ is the total number of spins in the sample, $%
\theta $ is the exponent of the initial increase in the magnetization, and d$%
_{eff}$ is the scaling dimension whose value may be different than d$_H$.
Also, $\beta $ and $\nu $ are the standard-usually defined in equilibrium-order 
parameter and correlation length critical exponents, respectively. The
scaling function behaves as $F(x)\sim 1$ for $x\rightarrow 0$ and $F(x)\sim
\frac 1x$ for $x\rightarrow \infty $. Extrapolating the results to $m_0=0,$
the exponent $\theta $ of the initial increase in the magnetization, which is a 
new \textit{nontrivial} critical exponent, can be obtained. We remark that 
$M^2(t)$ and $A(t)$ have to be measured starting from fully uncorrelated
configurations with strictly zero magnetization, i.e for $T=\infty $. One
should also expect that the generalized hyperscaling relationship given by $%
d_{eff}-\frac{2\beta }\nu =\gamma $ would hold for the Ising model on a
fractal substrate (see also equation (\ref{ec3})).

For a dynamic relaxation from a completely ordered state ($m_0=1$)
corresponding to the ground state at $T=0$, annealed to $T_c$, the
magnetization, the logarithmic derivative of the magnetization with respect
to the reduced temperature $\tau =\frac{T-T_c}{T_c}$, and the second-order
Binder cumulant should follow power laws in time, namely

\begin{equation}
M(t)\propto t^{-\frac \beta {\nu z}},  \label{ec5}
\end{equation}

\begin{equation}
V_\tau (t)=\partial _\tau (ln\;M(t,\tau ))\mid _{\tau =0}\propto t^{\frac
1{\nu z}},  \label{ec6}
\end{equation}

\begin{equation}
U(t)=\frac{M^2(t)}{\left[ M(t)\right] ^2}-1\propto t^{\frac{d_{eff}}z},
\label{ec7}
\end{equation}

\noindent respectively. For $T\neq T_c$, but within the critical region, the power-law
behavior is modified by a scaling function, which for the magnetization is
given by $M^{*}(t^{\frac 1{\nu z}}\tau )$. So, this fact can be used to
determine the critical temperature from the localization of the optimal
power-law behavior.

We have found that the dynamic evolution of the magnetization when the
system is annealed from $T=0$ to a higher temperature $T$ close to the
critical point is more sensitive to this final temperature than when the
system is quenched from T=$\infty $. Furthermore, due to the large initial
value of $M$ and its slow decrease upon annealing, the statistical
fluctuations are less prominent, and therefore it would be expected that the
topological effects may be easier to detect.

\subsection{Monte Carlo Simulations}

In order to study the effects of the topology of the fractal on the STD and
the convergent behavior of the critical exponents with the segmentation
step, we carried out Monte Carlo simulations for segmentation steps $k=4,5,$
and $6$ (system size L = 256, 1024, and 4096, respectively) using periodic
boundary conditions. Simulations started either from the complete ordered
state or from a disordered state with zero or a small magnetization. In the
last case the initial magnetization m$_0$ was settled by flipping, in a
random disordered configuration, a definite number of spins placed at
randomly chosen sites in order to get the desired value of $m_0$. The time
evolution of the system was updated by means of a Metropolis algorithm using
the Marsaglia-Zanan pseudorandom number generator\cite{Vattulainen}. The
time unit, defined as a Monte Carlo step (MCS), involves attempts to update $%
N$ randomly selected spins. The time evolution was followed, depending on
the initial state, from 10$^3$ up to $2 \times$ $10^5$ MCS. The magnetization,
the autocorrelation, and the second moment of the magnetization were averaged
over a number $n_s$ of samples with the same initial conditions but using
different configurations.

\section{Results and Discussion}

\subsection{Dynamic Evolution from the Ordered State}

\begin{figure}[tbp]
\includegraphics[height=9cm,width=8cm,angle=-90]{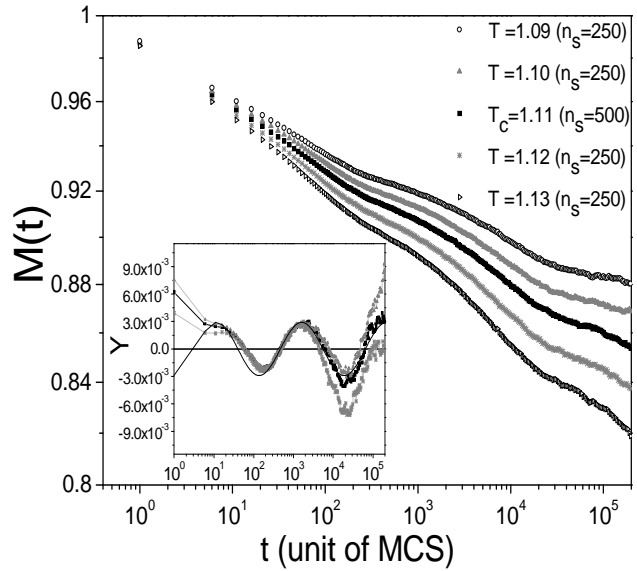}
\vskip 1.0 true cm
\caption{Log-log plots of magnetization versus time obtained for different
temperatures close to criticality, starting from ordered initial conditions 
($m_{0} = 1$) for the segmentation step k=5. The critical temperature 
$T_{c}(k=5) = 1.11(1)$ was determined by obtaining the best fit of $M(t)$ to
equation (\ref{ec8}). The inset shows the oscillating component of $M(t)$ 
$(Y=\frac{M(t)-C t^{-\frac{\beta}{\nu z}}}{C t^{-\frac{\beta}{\nu z}}})$
versus $t$ in log-scale for T=Tc, T=1.10 and T=1.12, which we fit with a
cosine function (solid line). More details in the text.}
\label{fig2}
\end{figure}

Figure \ref{fig2} shows the decay of the magnetization as a function of time
obtained at different annealing temperatures for the segmentation step k=5.
As can be observed, log-periodic oscillations are present in the time
evolution of the magnetization over the full range of time and for all
temperatures. We believe that the log-periodic structure exhibited by the
data is a typical feature of systems with Discrete Scale Invariance (DSI)%
\cite{Sornette}. So, in Section III C we first show that the topology of the
substrate (the Sierpinki carpet in our case) possesses this property and
subsequently, we formulate arguments linking the spatial characteristic
length of the fractal to the time characteristic period of the dynamic
behavior of the Ising system. Now, in order to account for our numerical
observations we assume that the power-law behavior expected for the time
evolution of the magnetization (equation \ref{ec5}) should be replaced by

\begin{figure}[tbp]
\includegraphics[height=9cm,width=8cm,angle=-90]{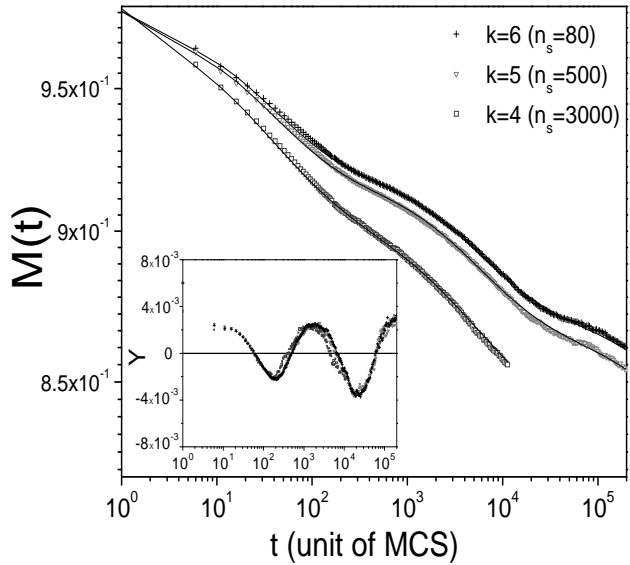} \vskip 1.0 true
cm
\caption{Log-log plots of magnetization versus time obtained at the critical
temperature starting from ordered initial conditions ($m_{0} = 1)$ for the
segmentation step k=4, 5, and 6. The corresponding fits are also shown by
mean full lines. The inset shows the shape of the oscillation of $M(t)$ 
$(Y=\frac{M(t)-C t^{-\frac{\beta}{\nu z}}}{C t^{-\frac{\beta}{\nu z}}})$ versus 
$t$ in log-scale.}
\label{fig3}
\end{figure}

\begin{equation}
M(t)\propto t^{-\frac \beta {\nu z}}\left( 1+A\,\cos \left( \frac{2\pi }{%
\log (P)}\log (t)+\phi \right) \right),  \label{ec8}
\end{equation}

\noindent  where log($P$) is a logarithmic period, while $A$ and $%
\phi $ are the amplitude and the phase of the oscillation, respectively. An
similar expression to equation(\ref{ec8}) has been used by other authors
to describe the time dependence of systems that exhibit a DSI in their dynamic
behavior, such as the energy release on the approach of impending rupture,
earthquakes\cite{Sornette}, and biological systems (proteins)\cite{Shen}. In
order to study the critical regime we settle down the temperature range that
exhibits the smallest deviations from a power-law behavior, namely 1.09 $\leq
T_c\leq $ 1.13. Subsequently, we determine the critical temperature by
finding the smallest deviation of the data from equation (\ref{ec8}). For 
$k=5$, this study yields $T_c=1.11(1)$, where the error bar was assessed by
considering the closest pair of temperatures that present noticeable but
small deviations. The inset of Figure \ref{fig2} shows the oscillatory
component of $M(t)$ that is nicely fitted by the cosine function for $T=1.11$,
and also shows the departures from the cosine function for $T=1.10$ and $%
T=1.12$. Logarithmic periodic oscillations are also observed for the
segmentation steps $k=4$ and $6$, and the procedure previously described was
also used to determine the corresponding critical temperatures, as reported
in Table \ref{tab1} (2nd column). Figure \ref{fig3} shows the magnetization
decay at the critical temperature for the segmentation steps 4, 5, and 6.
The fits were carried out after disregarding an initial time interval $%
t_{min}$ = 30MCS, and the obtained parameters are reported in Table \ref
{tab1}. For $k=4$, in order to assure that the nonequilibrium correlation
length remains smaller than the system size, the considered time interval
was 30-10$^4$MCS. As observed in the inset of Figure \ref{fig3} the
logarithmic periodic oscillations of the magnetization decay have similar
shape for k=5 and 6, but they are slightly different for k= 4. The critical
temperature $T_c(k=6)$= 1.10(1) is in relatively good agreement with
determinations performed by means of Monte Carlo simulations in equilibrium
and obtained by applying finite-size scaling, namely $T_c<$1.178\cite
{Monceau04} and $T_c$ =1.077(3)\cite{Carmona}. Furthermore, a careful
inspection of the data reveals a systematic but convergent decrease in both $%
T_c(k)$ and $\beta /\nu z$ (Table \ref{tab1}, 3rd. column) when the
segmentation step is increased, suggesting that our results could be taken
as upper bounds. The exponent $\beta /\nu z$ for k = 6 is notably smaller
than both the accepted value for the Ising model in $d=2$, given by $\beta
/\nu z=0.0577(3)$ \cite{Zheng00}, and our estimation for the SC(3,1), given
by $\beta /\nu z=0.03412(7)$ \cite{Bab05}. In this way at the critical point
a larger deviation from the translational symmetry, i. e. an increase in the
lacunarity, makes the decay of the long-range order slow down.

\begin{table}[tbp] \centering%
\caption{Critical temperatures and parameters obtained by fitting the time
dependence of the magnetization according to equation(\ref{ec8}): ratio of  critical exponents 
$\beta /\nu z$, and logarithmic period (log(P)), amplitude (A) and phase ($\phi $) of the
oscillation. Data obtained by starting the simulations from an ordered initial
state and for segmentation steps k= 4, 5, and 6.}\label{tab1} 
\begin{tabular}{cccccc}
\hline\hline
$k$ & $\mathbf{T}_c(k)$ & $\frac \beta {\nu z}$ & $A$ & $\log (P)$ & $\phi $
\\ \hline\hline
4 & 1.14(1) & 0.0145(1) & 0.0023(2) & 2.11(7) & $\pi $ \\ 
5 & 1.11(1) & 0.0116(1) & 0.0029(1) & 2.15(1) & $\pi $ \\ 
6 & 1.10(1) & 0.0110(1) & 0.0029(1) & 2.16(1) & $\pi $ \\ \hline\hline
\end{tabular}
\end{table}%

\begin{figure}[tbp]
\includegraphics[height=9cm,width=8cm,angle=-90]{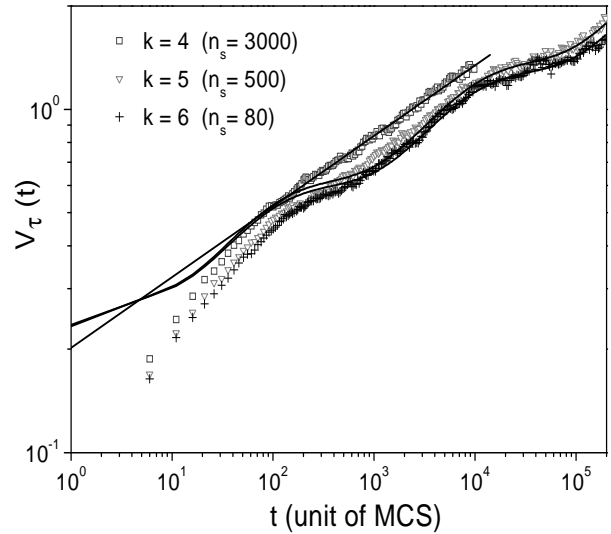} \vskip 1.0 true
cm
\caption{Log-log plot of the logarithmic derivative of magnetization versus
time obtained at $T_c$ by starting from the ordered initial conditions 
($m_{0} = 1)$. Results obtained for three different segmentation steps as
listed in the figure. The full lines correspond to fits obtained for $t >
100 $ MCS, according to equation (\ref{ec9}) for $k=5$ and 6, and to
equation(\ref{ec6})for $k=4$, respectively. See details in the text.}
\label{fig4}
\end{figure}

\begin{figure}[tbp]
\includegraphics[height=9.1cm,width=8cm,angle=-90]{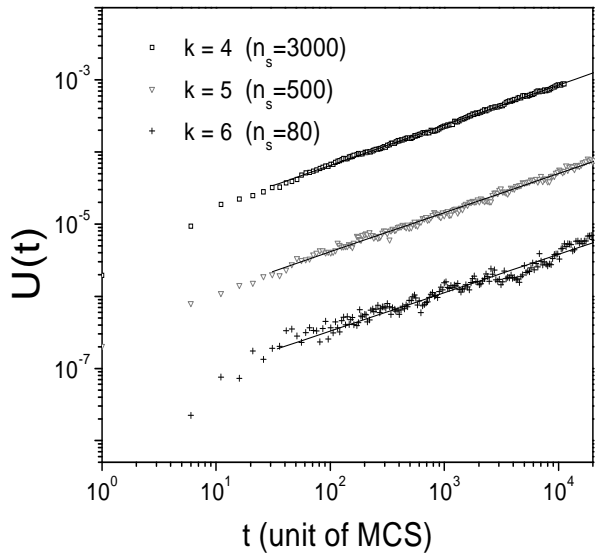} \vskip 1.0 true
cm
\caption{Log-log plot of Binder's cumulant versus time obtained at
criticality, starting from ordered initial conditions ($m_{0} = 1)$, for
k=4, 5, and 6. The full lines correspond to the best fits obtained for $t >
30 $ MCS.}
\label{fig5}
\end{figure}

\smallskip The logarithmic derivative of the magnetization with respect to $%
\tau $ is evaluated by taking the difference between the values of $M(t)$ at
two temperatures close to the critical one ($T_c\pm 0.01)$.{\Large \ }The
results of this calculation are shown in Figure \ref{fig4}, where the
logarithmic periodic oscillations can be clearly observed for the
segmentation step $k=6.$ For $k=5$ the oscillations are also suggested by
the behavior obtained for $t>10^3$MCS. In order to obtain the exponent $%
1/\nu z,$\ we propose the same correction for the power law as that
applied to the magnetization and given in equation (\ref{ec8}), namely

\begin{equation}
V_\tau (t)\propto t^{\frac 1{\nu z}}\left( 1+B\,\cos \left( \frac{2\pi }{%
\log (P)}\log (t)+\Phi \right) \right) ,  \label{ec9}
\end{equation}
where $B$ and $\Phi $ are the amplitude and the phase of the oscillations,
respectively. The fits were performed within the range 10$^2$\ to $2\times$ 10$
^5$MCS giving a value for log(P) = 2.0(2) and the exponents $1/\nu z$\ listed
in Table \ref{tab2}. For k = 4, the presence of oscillations cannot be
detected within the time range considered, i.e. from 10$^2$ to 10$^4$MCS.
In order to give a crude estimation for the critical exponent, we
fitted the data with equation (\ref{ec6}) (see Table \ref{tab2}). Figure \ref
{fig5} shows the determination of the exponent $\frac{d_{eff}}z$ (see Table 
\ref{tab2}) from the time dependence of the Binder cumulant, according to
equation (\ref{ec7}). The trend of the data for the exponents $1/\nu z$ and $%
\frac{d_{eff}}z$, namely a convergent decrease when $k$ is increased, is
consistent with the previous observations reported for SC(3,1)\cite
{Bab05,Pruessner} and strongly suggests that our results should be taken as
reliable upper bounds.

From these results the order parameter critical exponent $\beta $ can be
obtained, which appears to be less sensitive to the change of the
segmentation step than other exponents shown in Table \ref{tab2}. It is
worth mentioning that our best estimation, given by $\beta =0.068(3)$, is
significantly smaller than the values corresponding to the Ising model in $%
d=2$ ($\beta =0.125$(exact)) and in d$_H\cong 1.8927$ on SC(3,1)\ ($\beta
=0.121(5)$).

\begin{table}[tbp] \centering%
\caption{Critical exponents determined from the time dependence of
the magnetization (2nd column, equation(\ref{ec8})), the Binder
cumulant (3rd column, equation(\ref{ec7})) and the logarithmic
derivative of magnetization (4th column, equation (\ref{ec9})). Also, * corresponds to
a crude estimation performed by using equation (\ref{ec6}). Data obtained by
starting the simulations from an ordered initial state and for the k=4, 5, and 6 
segmentation steps of the fractal.}\label{tab2} 
\begin{tabular}{ccccc}
\hline\hline
$k$ & $\frac \beta {\nu z}$ & $\frac{d_{eff}}z$ & $\frac 1{\nu z}$ & $%
\mathbf{\beta }$ \\ \hline\hline
4 & 0.0145(1) & 0.556(3) & 0.209(2)* & 0.069(1) \\ 
5 & 0.0116(1) & 0.54(1) & 0.167(8) & 0.069(3) \\ 
6 & 0.0110(1) & 0.54(3) & 0.162(8) & 0.068(3) \\ \hline\hline
\end{tabular}
\end{table}%

\subsection{Dynamic Evolution from the Disordered State}

The dynamic evolution after quenching the system to $T_c$ when the
simulations are started from the disordered state presents a weak dependence
on temperature. This shortcoming hinders an independent estimation of $T_c$
based on these measurements. However, by using the value of $T_c$ obtained
by means of simulations started from the ordered state, it is possible to
obtain an independent evaluation of the critical exponents.

\begin{figure}[tbp]
\includegraphics[height=9cm,width=8cm,angle=-90]{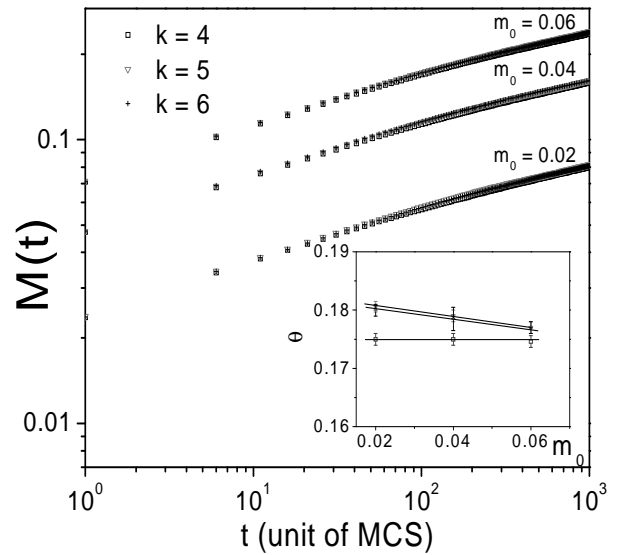} \vskip 1.0 true
cm
\caption{Log-log plots of magnetization versus time obtained at critical
temperature, starting from disordered initial conditions slightly modified
to obtain different values of the initial magnetization $m_{0}$. Data
corresponding to $k = 4 (n_{s}=50000), 5 (n_{s}=5000)$ and $6 (n_{s}=500)$,
and different values of $m_{0}$, which are also indicated in the figure. The
inset shows the dependence of $\theta$ on the initial magnetization $m_{0}$
that allowed us to extrapolate the exponent $\theta(m_{0} \rightarrow 0)$
for each $k$.}
\label{fig6}
\end{figure}

Figure \ref{fig6} shows the initial increase in magnetization observed for
the segmentation steps 4, 5, and 6, obtained for different values of the
small initial magnetization ($m_0=$ 0.02, 0.04, and 0.06). The data exhibit a
weak dependence on the segmentation step as can be deduced from the
overlapping of the curves. Within the time regime considered, the
magnetization always increases and the data can be fitted to a power law, as
expected from equation (\ref{ec2}). Nevertheless, a soft curvature of the
data can be observed for larger times due to the fact that $m_0$ is finite
and the power law is actually expected to hold in the $m_0\rightarrow 0$
limit. So, in order to determine the critical exponent we performed a fit of
the data within the time interval 30-100MCS. As can be observed in the
inset of Figure \ref{fig6}, the exponents show a weak dependence on $m_0$. Then, the
exponent $\theta $ was evaluated by a linear extrapolation to $m_0=0$. So,
according to our results, listed in Table \ref{tab3}, the exponent $\theta
=0.181(2)$ for the SC(4,2) fractal appears to be the same as that for SC(3,1) ($%
\theta =0.1815(6)$) and slightly smaller than the value for the Ising model
in $d=2$, given by $\theta =0.191(3)$ \cite{Zheng00}.

\begin{figure}[tbp]
\includegraphics[height=9cm,width=8cm,angle=-90]{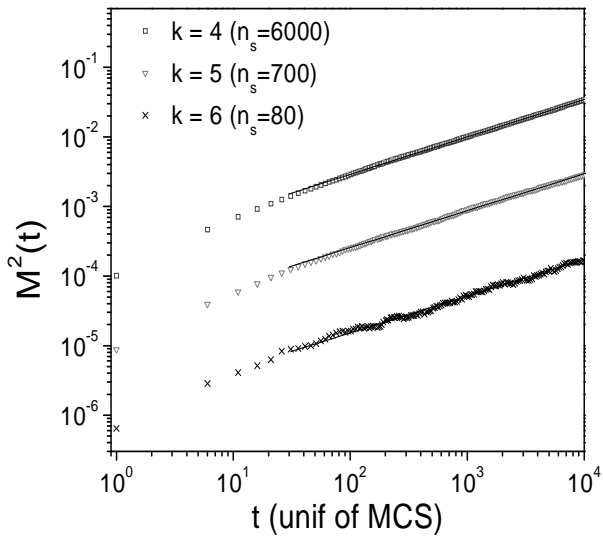} \vskip 1.0 true
cm
\caption{Log-log plots of the second moment of magnetization versus time,
obtained at the critical temperature, starting from disordered initial
conditions with $m_{0} = 0$. Data obtained by taking $k=4, 5,$ and $6$. The
full lines indicate the fit of the data according to equation (\ref{ec3})
for $t>30$.}
\label{fig7}
\end{figure}

\smallskip Figure \ref{fig7} shows the time evolution of the second moment
of the magnetization and the corresponding fit obtained, within the interval
30-10$^4$ MCS, according to equation (\ref{ec4}). The obtained values are
also listed in Table \ref{tab3}, which are within the error bars independent
of the segmentation step. The exponent $\gamma $ of the susceptibility can
be estimated assuming that hyperscaling law $d_{eff}=\frac{2\beta }\nu
+\frac \gamma \nu $ holds, and by combining the results corresponding to
different initial conditions, i.e. using the exponents of the second moment
and of the logarithmic derivative of the magnetization. The obtained values
(see Table \ref{tab3}) show a convergent increase with $k$, giving $\gamma
=3.3(2)$ for $k=6$. This value is significantly larger than those obtained
for the Ising model in $d=2$ and in SC(3,1), namely $\gamma =1.75$ (exact)
and $\gamma (k=6)=2.22(2)$, respectively. However, it is considerable
smaller than the values reported from the simulations performed in
equilibrium and obtained by using finite size scaling for the same
segmentation step, which are $\gamma \geq 5.39$\cite{Carmona,Monceau01}.

On the other hand, the decay of the autocorrelation function (see Figure 8)
slightly depends on the segmentation step. The fits were carried out after
disregarding an initial time interval of $t_{min}$ = 30MCS, since after that
the power-law behavior expected from equation (\ref{ec4}) is observed.
The obtained exponents $\lambda =d_{eff}/z-\theta $ are also reported in
Table \ref{tab3}. It is worth mentioning that by inserting the value of $%
\theta $ already determined from the initial increase in the magnetization
(see Figure \ref{fig6}) in the exponent of the autocorrelation function, one
can also calculate $d_{eff}/z$, as listed in the 4th column of Table \ref
{tab3}. The obtained results are in full agreement with independent
determinations performed by fitting the time dependence of the Binder
cumulant starting simulations from ordered states, see Table \ref{tab2}.

\begin{figure}[tbp]
\includegraphics[height=8.5cm,width=7cm,angle=-90]{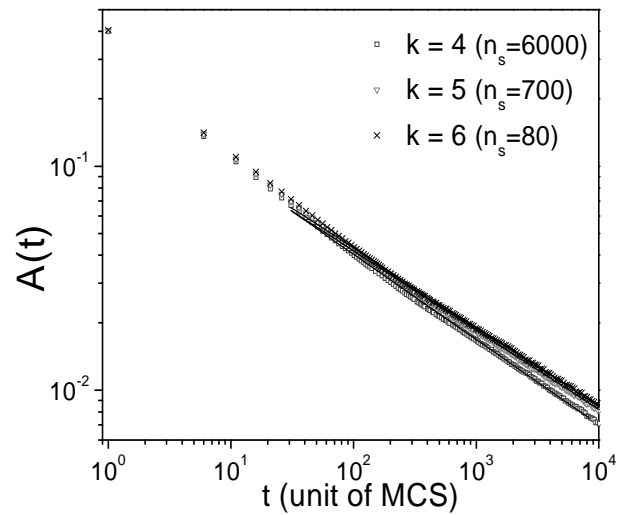} \vskip 1.0 true
cm
\caption{Log-log plots of autocorrelation versus time, obtained at critical
temperature, starting from disordered initial conditions with $m_{0}=0$ and
by taking $k =4, 5 $ and $6$. The full lines indicate the fit of the data
according to equation (\ref{ec4}) for $t>30$.}
\label{fig8}
\end{figure}

Also, by inserting the exponent $d_{eff}/z$ in the expression of the exponent
of the second moment of the magnetization, given by $d_{eff}/z-2\beta /\nu z$,
one can obtain an additional estimation of $\beta /\nu z$, as listed in
the 5th column of Table \ref{tab3}. The values are smaller than those obtained
for the magnetization decay corresponding to the evolution from the ordered
state (see the 2nd column of Table \ref{tab2}). We attribute this difference
to the small value of $\beta /\nu z$ relative to both $d_{eff}/z$ and $%
\left( \frac{d_{eff}}z-\frac{2\beta }{\nu z}\right)$. In this way, a small
relative error for these exponents induces a large error in $\beta /\nu z$,
as shown in Table \ref{tab3}.

\begin{table}[tbp] \centering%
\caption{Critical exponents determined from the dynamic behavior of
the second moment of the magnetization (2nd column, see equation(\ref{ec3})), 
the autocorrelation (3rd column, see equation (\ref{ec4})) and the initial
increase in the magnetization (4th column, see equation(\ref{ec2})). Data
obtained by starting the simulations from disordered initial states and for $k=4,5,6$.
From these exponents the  values of $d_{eff}/z$, $\frac \beta {\nu z}$, and $\gamma$ 
are estimated and are listed in columns 5 and 6 respectively. }\label%
{tab3} 
\begin{tabular}{ccccccc}
\hline\hline
$k$ & $\frac{d_{eff}}z\mathbf{-}\frac{2\beta }{\nu z}$ & $\frac{d_{eff}}z%
\mathbf{-\theta }$ & $\theta $ & $\frac{d_{eff}}z$ & $\frac \beta {\nu z}$ & 
$\gamma $ \\ \hline\hline
4 & 0.532(5) & 0.389(8) & 0.175(2) & 0.564.(8) & 0.016(9) & 2.55(3) \\ 
5 & 0.531(4) & 0.360(8) & 0.182(2) & 0.542(8) & 0.006(9) & 3.2(2) \\ 
6 & 0.529(5) & 0.360(9) & 0.181(2) & 0.541(9) & 0.006(10) & 3.3(2) \\ 
\hline\hline
\end{tabular}
\end{table}%

\subsection{Discrete Scale Invariance in Space and Time}

\smallskip Discrete Scale Invariance (DSI) has extensively been discussed by
Sornette\cite{Sornette}. It is a weak kind of scale invariance according to
which an observable $O(x)$, which is a function of a control parameter $x$,
obeys the scaling law 
\begin{equation}
O(x)=\mu (b)O(bx),  \label{ecdsi}
\end{equation}
under the change $x\rightarrow bx$. Here $b$ is no longer an arbitrary real
number as in the case of \textit{continuous} scale invariance but it can only
take specific discrete values, which in general have the form: $%
b_n=(b_1)^n$, where $b_1$ is the fundamental scaling ratio.

It is easy to show that if an observable $O(x)$ satisfies equation (\ref
{ecdsi}) for an arbitrary $b$ it necessary has to obey a power law of the
type $O(x)=Cx^\alpha $, where $\alpha$ is an exponent. But in the case of
DSI the solution of equation (\ref{ecdsi}) in general has the form

\begin{equation}
O(x)=x^\alpha F\left( \frac{\log (x)}{\log (b_1)}\right),  \label{ecdsi2}
\end{equation}
where $F$ is a periodic function of period one.

\begin{figure}[tbp]
\includegraphics[height=10cm,width=7cm,angle=-90]{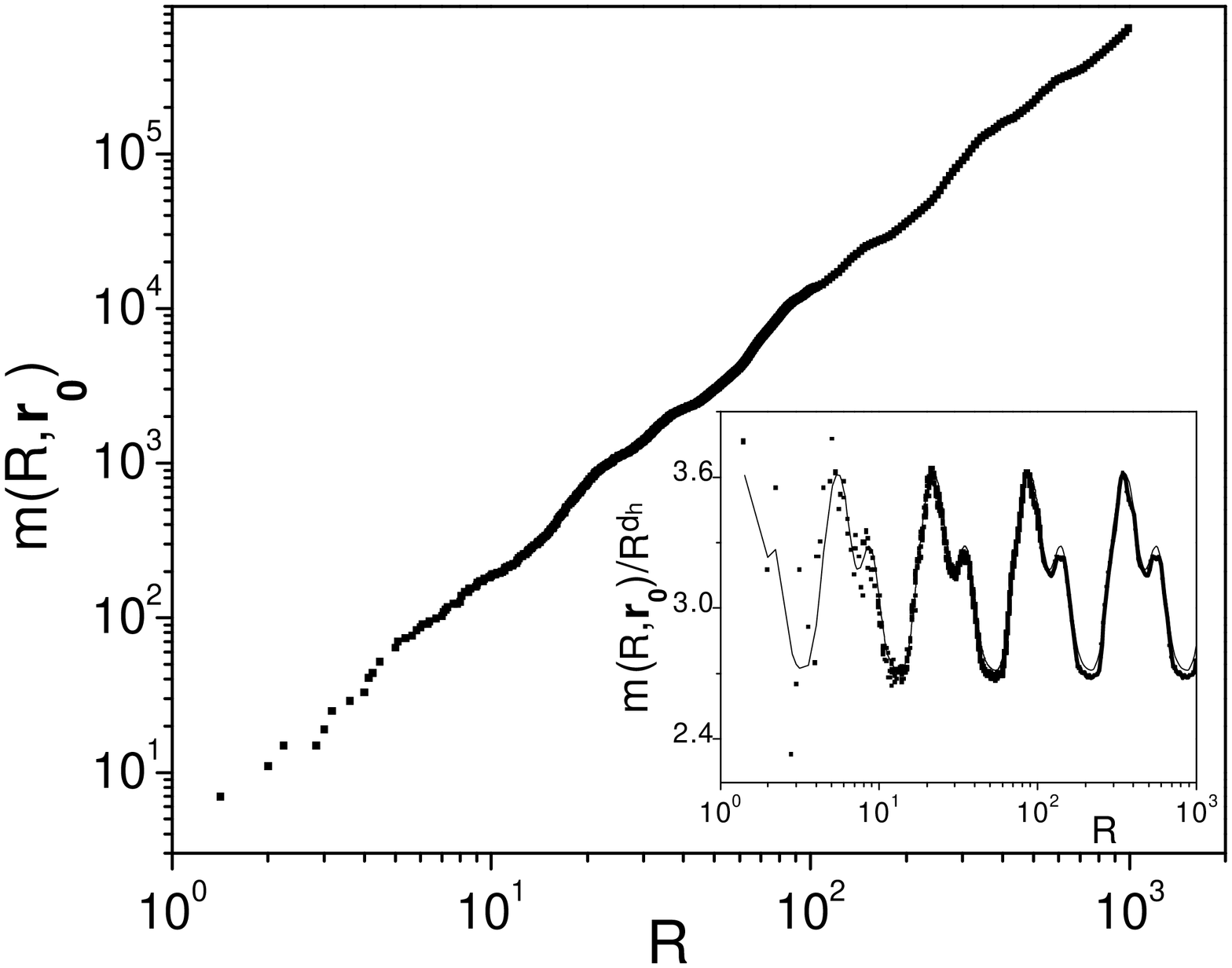} 
\vskip 1.0 true cm
\caption{Log-log plot of the mass $m(R,\mathbf{r_0})$ as a function of R,
for the segmetation step $k=5$ of the SC(4,2). The origin $\mathbf{r_0}$ has
been chosen at one of the fractal corners. The inset shows the logarithmic
oscillatory part, $P(\log(R)/ \log(b_1))= m(R,\mathbf{r_0})/R^{d_H})$, and a
fit with the first four terms of the Fourier expansion. For the fundamental
scaling ratio we have obtained $b_1=4.01(2)$.}
\label{fig9}
\end{figure}

To illustrate that the SC(4,2) leads up to DSI, let us calculate the
dependence of the mass for this fractal on the distance $R$ to an arbitrary
position $\mathbf{r_0}$ given by 
\begin{equation}
m(R,\mathbf{r_0})=\int_0^R\int_0^{2\pi }\rho (\mathbf{r-r_0})rd\theta
dr;~~~\rho (\mathbf{r})=\sum_{i=1}^N\delta (\mathbf{r-r_i})  \label{ecdsi3}
\end{equation}
where $i$ runs over all the sites in the fractal and $\mathbf{r_i}$ is the
position of the i-th site. Note that, in principle, the mass depends on the
point $\mathbf{r_0}$ we are choosing as coordinate origin. Figure \ref{fig9}
shows a log-log plot of the dependence of $m(R,\mathbf{r_0})$ on $R$, where $%
\mathbf{r_0}$ is taken as one of the corners of the fractal. As can be seen
in the inset, the mass oscillating component adjusts quite well to the behavior
expected from equation (\ref{ecdsi2}). The value for $b_1=4.01(2)$ obtained
by the fit with four terms in the Fourier expansion coincides with the
linear size of the generating cell, which of course is the fundamental
scaling ratio. By choosing different positions $\mathbf{r_0}$, the
dependence of the mass on $R$ looks different than the one shown in Figure 
\ref{fig9}. However, when we compute $m(R)$, the averaged mass function over
all the sites $\mathbf{r_0}$ in the lattice, we have find that it is still
possible to factorize a periodic part $F(\log (R)/\log (b_1))$ with $%
b_1\approx 4$. Therefore, we conclude that the mass, an exclusively
topological property of the fractal, presents a logarithmic oscillating
behavior, this a being signature of the DSI of the fractal.

We would like to remark that the fitted value of the fundamental scaling
ratio $b_1=4.01(2)$ for $k=5$ obtained from Figure (\ref{fig9}) is in
excellent agreement with the exact value given by $b_1=4$. This result
suggests a fast convergency of the logarithmic periodic behavior towards the
value corresponding to the mathematical fractal and gives us strong
confidence in the values of the characteristic time evaluated by fitting the
logarithmic periodic dynamics for $k=6$ (see Table (\ref{tab1})).

The observed logarithmic periodic oscillation in the decay of the
magnetization (Figures (\ref{fig2}) and (\ref{fig3})) leads us to abandon the
standard power-law decay of the form $M(t)\propto t^{-\frac \beta {\nu z}}$
(equation (\ref{ec5})) and propose a new Ansatz based on equation (\ref
{ecdsi2}), namely the typical power law modulated by a logarithmic periodic
function (see equation (\ref{ec8})). Of course, by means of this assumption
we implicitly recognize that the DSI intrinsic of the susbtrate is somewhat
capable of influencing the dynamic behavior of the overlaying ferromagnet
yielding to a DSI in the time scale. Following this line of reasoning one
has to admit that the fundamental scaling ratio between lengh scales 
 ($b_1$) of the fractal has to translate into an intrinsic ratio between time
scales for the oscillations given by the period $P$. It is well known that
for a critical system evolving towards equilibrium, characteristic spatial and 
temporal scales are linked
through the dynamic exponent $z$, i.e the development of the correlations
up to a length $\xi $ scales as $\xi \propto t^{1/z}$. Therefore, going one
step further in our speculation, we propose that the scaling ratios are also
linked according to

\begin{equation}
b_{1}=P^{\frac{1}{z}},  \label{ec12}
\end{equation}

In Appendix A we demostrate under which assumptions this relation holds, but
let us now show that equation (\ref{ec12}) is consistent with our
measurements. In fact, equation (\ref{ec12}) implies $z=\frac{log(P)}{%
log(b_1)}$. Then, from Table \ref{tab1} we take $log(P)=2.16$ yielding $%
z\approx 3.6$. Also, the logarithmic derivative of the magnetization yields $%
log(P)=2$, so one has $z\approx 3.3$. These rough estimations of the
dynamic exponent can be compared to the accepted value for the 2D-Ising
model, namely z=2.165 \cite{Zheng00} and our previous estimation of $z=2.55$%
\cite{Bab05} for the SC(3,1). So, all these results are consistent with an
increasing trend of $z$ when the fractal dimension of the substrate
decreases, anticipating the occurrence of a dramatic increase in the slowing
down, characteristic of the second-order phase transitions, in low
dimensionality. Furthermore, independent measurements (see Tables \ref{tab2}
and \ref{tab3}) yield to $\frac{d_eff}z\approx 0.54$, so one has $%
d_{eff}\approx 1.9$ ($z\approx 3.6$) and $d_{eff}\approx 1.8$ ($z\approx 3.3$%
), i.e. two figures in reasonable agreement with the fractal dimension of
the SC(4,2) given by $d_H\cong 1.7925$. On the other hand, based on both
the measured exponent $\frac 1{\nu z}=0.162(8)$ and the crude estimation of $%
z$, we could give a very rough estimation of the correlation length
exponent, which should be of the order of $\nu \approx 1.8$, also in
agreement with the trend shown for the 2D-Ising model ($\nu =1$) and SC(3,1)
($\nu \approx 1.39$).

\section{Conclusions}

We have studied the nonequilibrium critical dynamics in the short-time
regime of the Ising ferromagnet embedded in a fractal substrate, namely the
SC(4,2). The influence of the topology of the fractal on the dynamic
evolution of some physical observables is clearly identified through at
least two effects: i) The dependence of the short-time regime on the
segmentation step; and ii) the occurrence of logarithmic periodic
oscillations superimposed to the power-law behavior observed in both the
decay of the magnetization and its logarithmic derivative with respect to
the reduced temperature, when the system is annealed from $T=0$ to $T_c$.
We propose that these oscillations are a consequence of the DSI of the fractal substrate.

In order to describe the decay of the magnetization upon annealing to $T_c$
we proposed an Ansatz that involves the standard power-law behavior but now
modulated by a logarithmic periodic oscillatory function (see equation (\ref
{ec8})). This kind of function has also been used to describe the behavior
of systems that exhibit a DSI in its dynamic behavior\cite{Sornette,Shen}.
By fitting the data with equation (\ref{ec8}) we determine $T_c=1.10(1)$%
, a figure that is in relatively good agreement with the values obtained 
by other authors by
performing equilibrium measurements and by applying a finite-size scaling
approach\cite{Carmona,Monceau04}.

The exponent $\theta =1.81(2)$ of the initial increase in the magnetization,
determined for the segmentation step $k=6$, appears to be the same as that for
the SC(3,1) ($\theta =0.1815(6)$) but it is only slightly smaller than the
accepted value for the Ising model in $d=2$ ($\theta =0.191(3)$), suggesting
that this exponent is not significantly affected by the dimensionality of
the substrate. It is worth mentioning that $\theta $ is related to $x_0$,
such that $x_0=z\theta +\frac \beta \nu $ and that the former exponent sets
the time scale for the initial increase in the magnetization through $%
t_{mic}\simeq m_0^{-\frac{x_0}z}$.

The second moment of the magnetization and the autocorrelation function
obtained by starting the dynamic evolution from the disordered initial state
follow the expected power-law behavior. This fact allows us to perform an
independent determination of the exponents $\frac{deff}z$ and $\frac \beta
{\nu z}$, which turn out to be self-consistent with the values obtained from
simulations starting from the ordered state.

We would like to remark that the critical exponents corresponding to all
observables measured show a convergent trend when the segmentation step is
increased, indicating that the largest segmentation used in this work ($k=6$)
would be considered a good approximation of the mathematical fractal.

Our estimation $\beta =0.068(3)$ indicates that the order parameter critical
exponent for the SC(4,2) is smaller than that corresponding to the Ising
model in $d=2$ ($\beta =1/8$), which is almost the same as that reported for
the case as the SC(3,1) ($\beta =0.121(9)$). On the other hand, our
estimation for the exponent of the susceptibility ($\gamma =3.3(2)$) is
significantly larger than those corresponding to the 2D-Ising magnet 
($\gamma =1.75)$) and the SC(3,1) for the same k ($\gamma =2.22(2)$).
Nevertheless, it is notably smaller than the values reported from equilibrium
simulations using finite-size scaling ($\geq 5.39$)\cite{Carmona,Monceau04}.

We hope that the present work represents an extensive attempt to not only
numerically characterize the relevant critical properties of the Ising model
on a fractal substrate, but also to give clear evidence that the DSI of the 
underlying fractal structure influences the dynamic evolution of an 
Ising magnet. Of course, we have discussed the difficulties one encounters when
dealing with these systems, but even more importantly our study allows us to
clearly identify and characterize the influence of the substrate topology
on the power-law behavior expected for the nonequilibrium critical dynamics
in the short-time regime. The DSI of the fractal shows up dramatically in
the dynamic behavior of the magnetization decay by causing the occurrence of
logarithmic periodic oscillations.

\section{Appendix A}

In this appendix we show that equation (\ref{ec12}) can be straightly proven
for a certain observable $M$, if one assumes:
\begin{itemize}
\item [(i)] $M(t)$ obeys a time DSI of the form:
\begin{equation}
M(t)=\mu(P_n)M(P_n t), ~~ P_n=P^{n} \label{app1}
\end{equation}
where P is the period and n is an integer.

\item [(ii)] The correlation lenght scales as 

\begin{equation}
\xi\propto t^{1/z}, \label{app2}
\end{equation}

\item [(iii)] The same observable $M$, as a function of the correlation
lenght, obeys a spatial DSI of the form:

\begin{equation}
\tilde{M}(\xi)=\tilde{\mu}(b_n)\tilde{M}(b_n \xi), ~~ b_n=b_1^{n} \label{app3}
\end{equation}

where $b_1$ is the fundamental scaling ratio between lenght 
scales and $\tilde{M}$ is using to denote: $\tilde{M}(\xi)=M[t(\xi)]$.
\end{itemize}

Assumption $(ii)$ implies $t=a \xi ^z$ for some constant $a$. Then by
replacing on the right-hand side of equation (\ref{app1}) one obtains
\begin{equation}
M(t)=\mu(P^n)M(P^n a \xi ^z)=
\mu(P^n)M[ a (P^{\frac{n}{z}}\xi) ^z],
\end{equation}
and writing $M(t)$ as $\tilde{M}(\xi)$ this equality becomes
\begin{equation}
\tilde{M}(\xi)=\mu(P^n) \tilde{M}(P^{\frac{n}{z}}\xi),
\end{equation}

The comparison with  equation (\ref{app3}) leads us to the the following equalities

\begin{equation}
P^{\frac{n}{z}} = b_1 ^n, ~~~ \tilde{\mu}(b_1^n)=\mu(P^n),
\end{equation}

and therefore, $P^{\frac{1}{z}} = b_1$.

\smallskip

Acknowledgements: This work was supported by CONICET, UNLP, ANPCyT and
Fundaci\'{o}n Antorchas (ARGENTINA). The A. von Humboldt Foundation
(Germany) is greatly acknowledged for the provision of valuable computer
equipment. Alberto Maltz and Ernesto Loscar are acknowledged for fruitful
discussions.

\vspace{1cm}


\begin{references}
\bibitem{Hao87}  L. Hao and Z. R.Yang, J. Phys. A, {\bf 19}, 1627 (1987).

\bibitem{Monceau04}  P. Monceau and, P. Hsiao, Physica A, {\bf 331}, 1
(2004).

\bibitem{Gefen83}  Y. Gefen, A. Aharony, and B. Mandelbrot, J. Phys.\ A, 
{\bf 16}, 1267 (1983).

\bibitem{Gefen80}  Y. Gefen, B. B. Mandelbrot, and A. Aharony, Phys. Rev.
Lett., {\bf 45}, 855 (1980).

\bibitem{Gefen84}  Y. Gefen, A. Aharony, and B. Mandelbrot, J. Phys. A, 
{\bf 17}, 1277 (1984).

\bibitem{Mandelbrot}  B. B. Mandelbrot, The fractal Geometry of Nature,
Freeman, San Francisco, 1982.

\bibitem{Monceau01}  P. Monceau and, M. Perreau, Phys. Rev. B, {\bf 63},
184420 (2001).

\bibitem{Pruessner}  G. Pruessner, D. Loison, and K. K. D. Schotte, Phys.\
Rev. B, {\bf 64}, 134414 (2001).

\bibitem{Janssen}  H. K. Janssen, B. Schaub, and B. Schmittmann, Z. Phys. B:
Condens. Matter, {\bf 73}, 539 (1989).

\bibitem{Zheng98}  B. Zheng, Int. Mod. Phys. B, {\bf 12}, 1419 (1998).

\bibitem{Bab05}  M.\ A. Bab, G. Fabricius, and E. V. Albano, Phys. Rev. E, 
{\bf 71}, 036139 (2005).

\bibitem{Sornette} D. Sornette, Phys. Rep. {\bf 297}, 239 (1998).

\bibitem{Lessa}  C. J. Lessa and R. F. S. Andrade, Phys. Rev. E, {\bf 62},
3083 (2000)

\bibitem{Shlesinger}  M. F. Shlesinger and B. J. West, Phys. Rev. Lett, {\bf 67},
2106 (1991).

\bibitem{Shen}  T. Y. Shen, K. Tai and J. A. McCammon, Phys. Rev. E, {\bf 63},
041902 (2001).

\bibitem{Isliker}  H. Isliker and L.\ Vlahos, Phys. Rev. E, {\bf 67}, 026413
(2003).

\bibitem{Monceau02}  P. Monceau and, P. Hsiao, Phys. Lett A, {\bf 300}, 687,
(2002).

\bibitem{Zheng99}  B. Zheng, M. Schulz and S. Trimper, Phys. Rev. Lett., 
{\bf 82}, 1891 (1999).{\bf \ }

\bibitem{Vattulainen}  I. Vattulainen, T. Ala-Nissila, and K. Kankaala,
Phys.\ Rev. E, {\bf 52}, 3205 (1995).

\bibitem{Carmona}  J. M. Carmona, U. M. B. Marconi, J. J. Ruiz-Lorenzo and
A. Taranc\'{o}n, Phys. Rev. B, {\bf 58}, 14387 (1998).

\bibitem{Zheng00} B. Zheng, Physica A, \textbf{283}, 80 (2000).
\end{references}
\end{document}